\documentclass[prl,aps,superscriptaddress,twocolumn,floats,nofootinbib,showpacs]{revtex4}
\usepackage{amsmath,amssymb,dsfont,citesort,graphicx}

\newcommand{\dx}{\partial_x}

\DeclareMathOperator{\sign}{sgn}

\newcommand{\PRL}[3]{Phys. Rev. Lett.~\textbf{#1}, #2 (#3)}

\newcommand{\PR}[3]{Phys. Rev.~\textbf{#1}, #2 (#3)}
\newcommand{\RMP}[3]{Rev. Mod. Phys.~\textbf{#1}, #2 (#3)}
\newcommand{\Science}[3]{Science~\textbf{#1}, #2 (#3)}

\newcommand{\JETP}[3]{Sov. Phys. JETP~\textbf{#1}, #2 (#3)}
\newcommand{\ZhETF}[3]{Zh. Eksp. Teor. Fiz.~\textbf{#1}, #2 (#3)}
\newcommand{\JMP}[3]{J. Math. Phys.~\textbf{#1}, #2 (#3)}

\newcommand{\etal}{\textit{et al.}}

\begin{document}

\title{Generating  dark solitons  by  single photons}

\author{M. Khodas}
\affiliation{William I. Fine Theoretical Physics Institute,
University of Minnesota, Minneapolis, MN 55455}
\affiliation{
School of Physics and Astronomy, University of Minnesota,
Minneapolis, MN 55455}
 \author{A. Kamenev}
\affiliation{ School of Physics and Astronomy, University of
Minnesota, Minneapolis, MN 55455}
\author{L. I. Glazman}
\affiliation{William I. Fine Theoretical Physics Institute,
University of Minnesota, Minneapolis, MN 55455}
\affiliation{
School of Physics and Astronomy, University of Minnesota,
Minneapolis, MN 55455}
 \affiliation{Department of Physics, Yale
University, P.O. Box 208120,
  New Haven, CT 06520-8120}

\begin{abstract}
  We show that dark solitons in 1D bose systems may be excited by
  resonant absorption of single quanta of an external ac field.  The
  field frequency $\omega$ should be slightly blue-detuned from
  $\varepsilon_s(\hbar q)/\hbar$, where $\varepsilon_s(\hbar q)$ is the
  energy of a soliton with momentum corresponding to the external
  field wavenumber $q$. We calculate the absorption cross-section and
  show that it has power-law dependence on the frequency detuning
  $\omega-\varepsilon_s(\hbar q)/\hbar$. This reflects the quantum nature of
  the absorption process and the orthogonality catastrophe phenomenon
  associated with it.
\end{abstract}

 \maketitle

The existence of dark solitons (DS) is probably the most spectacular
manifestation of the role played by {\em weak} inter-particle
interactions in 1D cold atomic gases \cite{SolitonObs}. Such
solitons are {\em macroscopically} large areas of partially, or even
completely depleted gas, which propagate without  dispersion.
It is natural to interpret these objects
as solutions of the {\em classical} Gross--Pitaevskii equation
\cite{SolitonBook,SolitonTheory}. Correspondingly, the means to
create DS, employed so far, required a macroscopic classical
perturbation applied to the atomic cloud. An example of the latter
is the phase imprinting technique \cite{Imprinting}, where a finite
fraction of the 1D cloud is subject to an external potential for a certain time.

\begin{figure}[h]
\includegraphics[width=0.7\columnwidth]{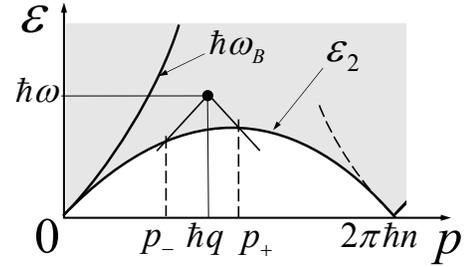}
\caption{Energy vs momentum plane for excitations. A photon is
  represented by a solid dot. The ``light cone'' with Bogoliubov
  velocity $v_B$ determines the range of possible momenta of generated
  dark solitons, $p_-<p_s<p_+$.  \label{fig:continuum}}
\end{figure}

Yet, there is a deep connection between DS and intrinsically {\em
quantum} nature of the gas. It was first appreciated by Kulish,
Manakov and Faddeev \cite{Kulish}, who noticed that the soliton's
dispersion relation coincides \cite{foot1} with the exact lower bound
of the {\em quantum many-body} spectrum of the Lieb-Liniger  model
\cite{Lieb}, Fig.~\ref{fig:continuum}.  The existence of such a bound,
known also as the Lieb II mode, was derived earlier using Bethe Ansatz
technique \cite{Lieb}. This observation means that the DS may be
considered as a classical approximation of very peculiar quantum
many-body eigenstates: those which possess the minimal possible energy
$\varepsilon_2$ at a given momentum $\hbar q$, see Fig.~\ref{fig:continuum}.

In this Letter we show that the underlying quantum nature of the
gas allows for a qualitatively different way of creating DS.
Namely, the soliton may be generated by the absorption of a {\em
single quantum} \cite{foot2} of an ac field slightly blue-detuned
from the resonance with the soliton energy $\hbar\omega\gtrsim
\varepsilon_2(\hbar q)$, where $\omega$ and $q$ are frequency and
wavenumber of the ac field. Notice, that the comparison between
the soliton's energy and the ac frequency does not appear in the
classical treatment at all.

The quantum efficiency of the discussed process is limited by the
orthogonality catastrophe~\cite{orthogonality}.  Indeed, a state of
the system after it is excited from the ground state by absorbtion of one quantum of the ac
field has a very small overlap with the eigenstates forming the
soliton. As a result, the quantum yield exhibits the very
characteristic power-law dependence on the energy excess above the
threshold: $\sim(\hbar\omega-\varepsilon_2)^{\mu_2}$.  The
corresponding exponent $\mu_2$ is a function of the wavenumber $q$
and the dimensionless interaction parameter $\gamma$.  For weak interactions, $\gamma\ll 1$, the exponent is
large, $\mu_2\sim\gamma^{-1/2}\gg 1$, reflecting the strong orthogonality.
The photon absorption and DS production, associated with it, are
higher for a moderate  interaction parameter, $\gamma \lesssim 1$, when
the DS density depletion is weaker.

To set the notation let us briefly remind what the localized
solutions of the non-linear Gross-Pitaevskii  equation are.
Quasiclassically, the condensate wave function obeys the following
equation \cite{review1999}:
\begin{align}                                                                       \label{GP}
i  \partial_t \Psi + \frac{ \hbar }{ 2 m }\, \partial_x^2 \Psi + c \left(
n  - |\Psi|^2 \right)\Psi = 0 \, ,
\end{align}
where $n=N/L$ is the average concentration and $L$ is the length of
the system with the periodic boundary conditions.
The interaction strength $c$ gives rise~\cite{Lieb} to the
dimensionless parameter $\gamma = c m /(\hbar n)$ whose smallness
$\gamma\ll 1$ is the criterion of the weakly interacting gas.

\begin{figure}[h]
\includegraphics[width=1.0\columnwidth]{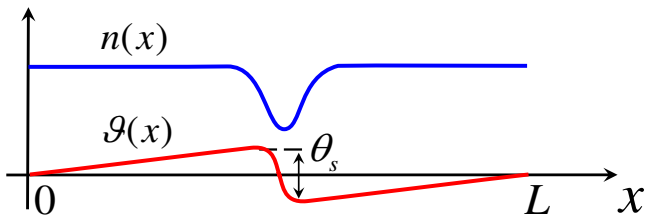}
\caption{(color online) Density $n(x)$ and phase $\vartheta(x)$
  profiles of a soliton in a system of length $L$, as given by
  Eq.~(\ref{solitonGP}). Note that the density perturbation is local,
  while the perturbation of phase is not.\label{fig:profile}}
\end{figure}

Soliton solutions of Eq.~\eqref{GP} have the form \cite{Tsuzuki1970}
\begin{align}
                                              \label{solitonGP}
\Psi_s =  \sqrt{n} \left[\cos \frac{ \theta_s }{ 2 } -
i \sin  \frac{ \theta_s }{ 2 }\, \tanh \left( \frac{ x - v_s
t}{l_s} \right) \right]e^{i{2x-L\over 2L}\theta_s} ,
\end{align}
where the only free parameter $\theta_s$ is the change of phase of
the condensate wave function $\Psi_s = \sqrt{n(x)}e^{i
\vartheta(x)}$ across the soliton. The condensate
density $n(x)$ reaches equilibrium value $n$ away from the soliton
(see Fig.~\ref{fig:profile}). For a given $\theta_s$ the soliton
velocity and its spatial extent are fixed and given by $v_s = v_B
\cos (\theta_s/2)$ and $ l_s = \hbar\left( m v_B \sin (\theta_s /
2) \right)^{-1}$, where $v_B$ denotes the Bogoliubov sound
velocity, which in the weakly interacting gas is given by $v_B =
\sqrt{\gamma}\,n\hbar /m $. The soliton,~Eq.~\eqref{solitonGP},
represents the density depletion, propagating without dispersion
with the velocity $v_s$. The number of particles pushed away from
the soliton core is
\begin{equation}\label{Ns}
N_s =  \frac{ 2K} { \pi}\, \sin { \theta_s \over 2 }\, ,
\end{equation}
where $K=\pi n\hbar /(m v_B)$ is the thermodynamic compressibility of
the gas.

%

In the presence of a soliton the  system acquires a non-zero
momentum $p_s$ and energy $\varepsilon_s$ which are expressed through
the phase shift $\theta_s$ as:
\begin{equation}\label{momentum}
  p_s = \hbar\, n
\left(\theta_s  - \sin \theta_s \right)\,; \quad \varepsilon_s =
{4\over 3}\, \hbar\, n v_B   \sin^3 {\theta_s \over 2}\, .
\end{equation}
These two relations  implicitly define the soliton dispersion
relation $\varepsilon_s=\varepsilon_s(p_s)$. One may check
that the velocity of the soliton satisfies  $v_s =
\partial \varepsilon_s / \partial p_s$, as expected for a particle. It is important to
stress that while the energy $\varepsilon_s$ is entirely associated
with the soliton core, the momentum $p_s$ is shared between the core
and the rest of the gas. More precisely, the soliton core carries
the momentum $-\hbar\,n\sin \theta_s$ (the negative sign here
corresponds to a hole in the density), whereas the momentum
$\hbar\,n\theta_s$ is uniformly spread to the rest of the gas. The
latter fact is easily seen from the phase $\vartheta(x)$ profile,
Fig.~\ref{fig:profile}, and the observation that the momentum
density of the condensate is given by $\hbar\,n\partial_x\vartheta$.

At small momenta $p_s\ll \pi\hbar n$, one finds $\theta_s\to 0$ and
thus $\varepsilon_s\approx v_B p_s$, cf. Eq.~(\ref{momentum}). In this
limit the soliton dispersion approaches the phonon branch
\cite{Pitaevskii} characterized by the Bogoliubov spectrum
$\omega_B(q)= (v_B q)\sqrt{1 + (\hbar q/2 m v_B )^2}$,
see Fig.~\ref{fig:continuum}. On the other hand, for
$p_s=\pi\hbar n$, the phase shift is $\theta_s=\pi$ so the soliton is
at rest: $v_s=0$.  In the latter case the entire momentum $p_s$ is
uniformly spread over the bulk of the gas.

Imagine now that the gas is subject to a weak  space and time
dependent external potential $V_0\cos (qx-\omega t)$. According to
the Fermi Golden rule, the system may absorb quanta of this field
if its many-body spectrum possesses excited states with the
momentum $\hbar q$ and energy $\hbar\omega$. As may be learned  from the
exactly solvable model  \cite{Lieb},  such states form  a
continuum whose energy is bound from below by the line $\varepsilon_2(\hbar q)$
(so called Lieb II mode). It was subsequently  noticed in
Ref.~\cite{Kulish} that in the limit of weak interactions
$\gamma\ll 1$ the Lieb II mode coincides with the soliton
dispersion relation, i.e. $\varepsilon_2(\hbar q)=\varepsilon_s(\hbar q)$. These
observations imply  that an absorption of a quantum blue-detuned
from the soliton energy $\hbar\omega\gtrsim \varepsilon_s(\hbar q)$ is (i)
possible and (ii) leads to  creation of the soliton. Our aim is to
study the scattering crossection of such processes.

To this end we notice that a photon absorption first creates a {\em
  virtual} state of the condensate with a {\em local} perturbation of
the condensate wavefunction. Since the photon carries momentum $\hbar
q$ and carries no extra particles, so does the initial local
perturbation. Subsequently this perturbation evolves according to the
equations of motion and eventually it must take a form of a
superposition of {\em real} excitations, i.e. conserving overall
energy $\hbar\omega$ in addition to the momentum $\hbar q$. We expect
that such a final state contains a soliton with momentum $p_s\approx
\hbar q$ and core energy $\varepsilon_s(p_s) < \hbar\omega$. The small
excess energy $\hbar\omega-\varepsilon_s>0$ and the momentum difference
$\hbar q-p_s$ are carried away by the small-amplitude sound waves
(phonons) with velocity $v_B$. This observation immediately
implies that $v_B |\hbar q-p_s|\leq (\hbar\omega-\varepsilon_s(p_s))$,
with the equality reached if all the sound waves are emitted in one
direction only. As a result, the range $p_- \leq  p_s  \leq p_+$
of possible soliton momenta $p_s$ is limited
\begin{align}                                                                       \label{interval}
 p_{\pm}(q,\omega) \approx \hbar
q \pm \frac{\hbar\omega - \varepsilon_s(\hbar q)}{v_B \pm v_s(\hbar q)
}\,; \quad\,\,\,\, p_+ - p_- \ll p_s\,.
\end{align}
The corresponding construction is shown in Fig.~\ref{fig:continuum}.

\begin{figure}[h]
\includegraphics[width=0.8\columnwidth]{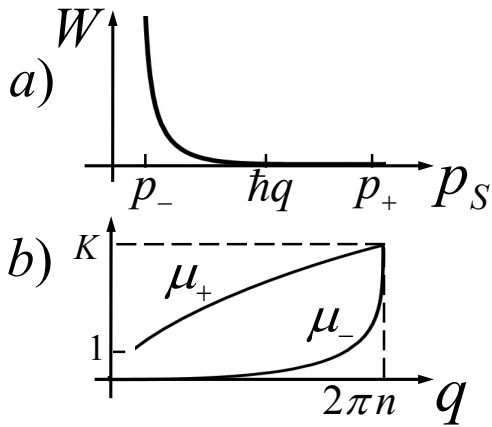}
\caption{ (a) Inelastic scattering crossection as
  a function of the soliton momentum $p_s$. (b) Momentum dependence of the
  exponents $\mu_{\pm}$  for $K=10$.  The
  exponent in Eq.~(\ref{eq:res}) is given by $\mu_2(q) = \mu_+(q) +
  \mu_-(q) - 1$.
\label{fig:crossection}}
\end{figure}

We found that the inelastic scattering crossection for absorbing a
photon with the  wavenumber $q$ and frequency $\omega$, while
creating the dark soliton with the momentum $p_s$ is given by
\begin{align}                                                                       \label{distribution}
W_{q,\omega}(p_s) \sim \frac{l_s}{\hbar v_B}\left[\frac{p_+  -
p_s}{\hbar/l_s} \right]^{\mu_+ - 1}
\left[\frac{p_s  -  p_-}{\hbar/l_s}\right]^{\mu_- - 1}\, .
\end{align}
It is characterized by the power-law dependencies on the
deviations of the soliton momentum from the upper and lower
kinematic boundaries $p_\pm(q,\omega)$. The corresponding
exponents $\mu_\pm$ are functions of the soliton parameter
$\theta_s=\theta_s(q)$ and the thermodynamic compressibility $K$,
\begin{equation}
\mu_{\pm}(q) = \frac{K}{4} \left(\frac{\theta_s}{\pi}  \pm \frac{
N_s}{ K}  \right)^2= \frac{K}{\pi^2} \left(\frac{\theta_s}{2} \pm
\, \sin \frac{ \theta_s}{ 2}  \right)^2. \label{exponents}
\end{equation}
In the last equality here we employed Eq.~(\ref{Ns}), and the
function $\theta_s(q)$ is implicitly defined by Eq.~(\ref{momentum})
with $p_s=\hbar q$. The calculated crossection as a function of the
soliton momentum $p_s$ for a photon with $q=\pi n$  is plotted in Fig.~\ref{fig:crossection}a.
We have also plotted the $q$--dependence of exponents $\mu_\pm(q)$
in Fig.~\ref{fig:crossection}b.

Equations \eqref{interval} -- \eqref{exponents} are the main result
of this paper. They provide the probability of the soliton
excitation with a specific momentum $p_s$ (and therefore specific
velocity $v_s=\partial \varepsilon_s/\partial p_s$). One may also be interested in the integral
probability of absorbing a photon with a given $q$ and
$\omega\gtrsim\varepsilon_s(\hbar q)/\hbar$, which results in creation
of a soliton with an unspecified momentum. This quantity is nothing
but the dynamic structure factor (DSF) $S(q,\omega)$ of the 1D Bose
gas. Integrating $W_{q,\omega}(p_s)$, Eq.~(\ref{distribution}), over
the soliton momenta $p_s$, one finds for DSF in an immediate
vicinity of the lower spectral boundary $\hbar \omega \gtrsim
\varepsilon_{2}(\hbar q)$
\begin{align}                                                                       \label{eq:res}
S(q,\omega) \sim {1\over v_B} \left[ \frac{ \hbar \omega -
\varepsilon_2(\hbar q)}{ \hbar v_B /l_s }\right]^{\mu_2(q)}\theta\left( \hbar \omega -
\varepsilon_2\right)\, .
\end{align}
Being multiplied by the intensity of the  radiation $V_0^2$ this
quantity gives a
number of solitons excited per unit time and per unit length of the
irradiated 1D gas. According to Eq.~(\ref{exponents}), the
wavenumber-dependent exponent $\mu_2(q) = \mu_+(q) + \mu_-(q) - 1$
is given by
\begin{align}                                                                      \label{eq:exp}
\mu_2(q) =
\frac{2K}{\pi^2}\left[\left(\frac{\theta_s}{2}\right)^2
                     +\left(\sin\frac{\theta_s}{2}\right)^2
                \right]-1\,,
\end{align}
%
The power law behavior of DSF near the lower spectral boundary
$\varepsilon_2(\hbar q)$ was suggested earlier in
Ref.~\cite{bosons2007}. However, Ref.~\cite{bosons2007} derived  functional dependence of $\mu_2(q)$
only in the limit of strong interactions, $\gamma\gg 1$.
Equation (\ref{eq:exp}) provides  the answer in the opposite limit of
weak scattering, $\gamma\ll 1$.

Below we outline the details of derivation. As was explained
earlier, the initial {\em local} perturbation of the condensate
carries the momentum of the absorbed photon $\hbar q$ and no excess
particles. After a short time, which may be estimated as
$\tau_s=l_s/v_B$, the soliton core is formed along with a bunch of
 outgoing sound
waves. The core, which is the density depletion, carries momentum
$-\hbar\, n\sin\theta_s$ and $-N_s$ particles. At times $t>\tau_s$
it propagates without dispersion and behaves as a free particle with
the energy $\varepsilon_s$. The remaining   momentum
$\hbar(q+n\sin\theta_s)=\hbar\,n\theta_s$, cf. Eq.~(\ref{momentum}),
and $N_s$ particles, localized on a scale $\sim l_s$ at
$t\sim\tau_s$ must be carried away and spread over the entire system
at $t\gg\tau_s$ by the linear {\em sound waves}. The latter are
conventionally described \cite{Haldane} by the linearized
hydrodynamic Hamiltonian
\begin{equation}                                             \label{Hhydro}
H_{sw} = \frac{\hbar \,v_B}{2\pi}\int\!dx \left[K^{-1} (\dx\hat{\varphi})^2 +
K(\dx\hat{\vartheta})^2\right]\, ,
\end{equation}
where $\hbar\, n\dx\hat\vartheta(x)$  and $\dx\hat\varphi/\pi$ are
operators of momentum  and  particle excess densities,
correspondingly. Their canonical commutation relation reads
$[\hat{\varphi}(x),\hat{\vartheta}(y)] =
i(\pi/2)\sign(x-y)$.

The injection of momentum $\hbar\, n\theta_s$ and $N_s$ particles
into the sound waves at a point $x$  is achieved by acting with operator
\begin{equation}
                                                   \label{solitonB}
\hat\psi_{sw}^{\dag}(x)= e^{i \frac{\theta_s}{ \pi } \hat{\varphi}(x) +
i N_s \hat{\vartheta}(x) }
\end{equation}
on the ground state of the Hamiltonian (\ref{Hhydro}). Indeed, the
operator $ e^{i \frac{\theta_s}{ \pi } \hat{\varphi}(x)}$ shifts $\hat
\vartheta(y)$ by $\theta_s$ at $x>y$, accommodating momentum
$\hbar\, n\theta_s$. Similarly, $e^{ i N_s \hat{\vartheta}(x) }$ shifts
$\hat\varphi(y)$ by $N_s$, at $x>y$ accommodating $N_s$ particles at the point $x$.
Since we are dealing with large shifts, we can
disregard non-commutativity of the operators, effectively adopting the
semiclassical approximation (see below).

The energy of a state created immediately upon the shift
Eq.~(\ref{solitonB}) is much higher than $\hbar\omega
(q)-\varepsilon_s(\hbar q)$. In full analogy with the instanton picture
of tunneling across a weak link in a Luttinger
liquid~\cite{fisher-lg}, the created virtual state evolves, eventually
reducing its energy to $\hbar\omega (q)-\varepsilon_s(\hbar q)$. To find
the photon absorption crossection resulting in a creation of a dark
soliton, one needs to evaluate the time evolution of the
density--density correlator of the excited sound waves,
\begin{equation}                                       \label{13}
G(x,t) =
\bigl\langle \hat\psi_{sw}(x,t)\hat\psi_{sw}^\dagger(0,0)\bigr\rangle .
\end{equation}
%
%
%
Its calculation uses Eqs.~\eqref{solitonB} and \eqref{Hhydro} and
follows the standard route of the bosonization theory
\cite{Haldane}, leading to
\begin{align}
                                                     \label{DW}
G(x,t)=
\left[1 + \frac{x-v_B t}{il_s} \right]^{-\mu_+}\!\!
\times \left[1 -\frac{x+v_B t}{il_s} \right]^{-\mu_-}\!\!.
\end{align}
Here the exponents are given by Eq.~(\ref{exponents}) and $l_s$
enters through the
short-time cutoff $\tau_s$ as $l_s=v_B\tau_s$. Finally the
absorption crossection is proportional to the Fourier transform of
the correlator:
\begin{equation}\label{WFourier}
W_{q,\omega}(p_s)\sim \mbox{Im}\!\int\!\! dx\, dt\, G(x,t)\, e^{i(\omega-\varepsilon_s/\hbar)t - (q-p_s/\hbar)x}\, ,
\end{equation}
where we took into account the momentum $p_s$ and energy
$\varepsilon_s(p_s)$ carried away by the soliton. To evaluate the
integral in Eq.~(\ref{WFourier}), we take into account that for
small energy excess $\hbar\omega-\varepsilon_s(\hbar
q)\ll\hbar\omega$ the range of the allowed soliton momenta $p_s$
is rather narrow, see Fig.~\ref{fig:continuum} and
Eq.~(\ref{interval}), and centered around the photon momentum
$\hbar q$. One may therefore expand the soliton energy as
$\varepsilon_s(p_s)\approx \varepsilon_s(\hbar q)+(p_s-\hbar q)v_s$.
Equations (\ref{interval}) and (\ref{distribution}) then follow
from the straightforward calculation.

The developed theory is applicable in the vicinity of the spectrum of
the Lieb II mode. The width of the corresponding region in the
$(q,\omega)$ plane, see Fig.~\ref{fig:continuum}, is determined by the
condition $\hbar\omega-\varepsilon_2(\hbar q) \lesssim \hbar/\tau_s$.
Also $\omega$ is restricted to be below the Bogoliubov mode $\omega_B$ and
its replica \cite{foot3}, shown by the dashed line in Fig.~\ref{fig:continuum}.
The soliton production rate reaches its maximum at
$\hbar\omega-\varepsilon_2(\hbar q) \sim \hbar/\tau_s$ and decreases at
higher frequency.

Additional restrictions are set by the applicability of semiclassical
approximation.  The semiclassical description of the Lieb II mode by
dark solitons fails at $p$ sufficiently close to $0$ or $2\pi\hbar n$.
This already can be seen from a comparison~\cite{unpub} of the true
spectrum of the Lieb II mode with the prediction of the semiclassical
Gross-Pitaevskii approach. At $\gamma\ll 1$ the two spectra
significantly differ from one another at $p\lesssim \hbar
n\gamma^{3/4}$. Equations (\ref{exponents}) and (\ref{eq:exp}) predict
monotonic decrease of the exponents $\mu_\pm$ and $\mu_2$ with the
decrease of $q$. For $q\ll \pi n$ they yield
\begin{equation}
\mu_2(q)\approx\mu_+(q)=\frac{K}{\pi^2}\left(\frac{6q}{
  n}\right)^{2/3}\!,\,\,\,\,\, \mu_-\propto q^2\ll\mu_+.
\label{small-q}
\end{equation}
This behavior is actually valid  in the interval $
n\gamma^{3/4}\ll q\ll\pi n$, the lower limit here being set by the
applicability of the semiclassical approximation. In this range
the exponents  $\mu_2, \mu_+\gg 1$, while there are no further restrictions
on $\mu_- > 0$.

Note that in the case of
strong interaction \cite{bosons2007} ($\gamma\gg 1$), the
asymptote at $q\to 0$ is $\mu_2(q)\propto q$. In this limit DS
degenerates into a single hole within  weakly-interacting {\em
Fermi sea}  in the  effective fermionic description \cite{Tonks}.
We expect the linearity must stay for the smallest $q$ at any
interaction strength, and thus expect a crossover from
$\mu_+,\mu_2\propto q^{2/3}$ to the linear behavior at $q\lesssim
n\gamma^{3/4}$. The similar crossover occurs  in the narrow
vicinity of the $q=2\pi n$ point, where the semiclassical soliton
description  also runs out of the applicability.

To conclude, we have shown that DS may be generated by absorption
of a single photon of an external ac field. Quantum efficiency of
such a process is maximized at slight, $\sim \hbar/\tau_s$, blue
detuning of the photon energy $\hbar\omega$ from the soliton
energy $\varepsilon_s(\hbar q)$. Within this range the absorption
probability behaves as the power law of the detuning, reflecting
the quantum orthogonality catastrophe phenomenon.

\begin{acknowledgments}
  We thank A. Abanov, J.-S. Caux, D. Gangardt,  D. Gutman, V.~Gurarie, and A.~Imambekov for numerous discussions.
  Research is supported by  NSF
  DMR-0749220 at Yale University, and by DOE
  (Grant DE-FG02-06ER46310) and A.P. Sloan foundation at the University of Minnesota.
\end{acknowledgments}


\begin{thebibliography}{99}
\bibitem{foot1} This statement is valid in the limit of weak
interactions.
\bibitem{foot3}
At $q\to 2\pi n$ Eq.~(\ref{eq:exp}) yields $\mu_2\approx 2K-1$, different from DSF exponent
$K-1$ established in the framework of the Luttinger liquid theory \cite{Haldane}.
The latter is applicable above the dashed line in Fig.~\ref{fig:continuum}.
\end{thebibliography}
\end{document}